\documentclass[showpacs,aps,prl,preprint,superscriptaddress,floatfix]{revtex4-1}
\usepackage{amsmath}
\usepackage{amssymb}
\usepackage{graphicx,color}
\usepackage{epstopdf}
\usepackage{ulem}

\begin{document}

\title{Conductance Discontinuity on the Surface of a Topological Insulator with Magnetic Electrodes}

\author{Xiaopeng Duan}
\affiliation{Department of Electrical and Computer Engineering, North Carolina State University, Raleigh, NC 27695, USA}

\author{Xi-Lai Li}
\affiliation{Department of Electrical and Computer Engineering, North Carolina
State University, Raleigh, NC 27695, USA}

\author{Yuriy G. Semenov}
\affiliation{Department of Electrical and Computer Engineering, North Carolina State University, Raleigh, NC 27695, USA}

\author{Ki Wook Kim}\email{kwk@ncsu.edu}
\affiliation{Department of Electrical and Computer Engineering, North Carolina State University, Raleigh, NC 27695, USA}
\affiliation{Department of Physics, North Carolina State University, Raleigh, North Carolina 27695, USA}

\begin{abstract}
Asymmetric electrical conductance is theoretically demonstrated on the surface of a topological insulator (TI) in the limit of infinitesimally small forward and reverse biases between two spin selective electrodes.  The discontinuous behavior relies on the spin-momentum interlocked nature of TI surface electrons together with the resulting imbalance in the coupling coefficients between the electrodes and TI surface states.  The analysis is based on a transmission matrix model that, in combination with a phenomenological treatment for the diffusive limit, accounts for both ballistic and scattered paths simultaneously.  With the estimated conductance asymmetry over a factor of 10, implementation in the ratchet-like applications and low-voltage rectification circuits appears practicable.
\end{abstract}

\maketitle

At the very dawn of semiconductor electronics, the two-terminal rectifier or diode was known as the device that breaks the conductance invariance with respect to the electric current (or bias) reversal. The significance of this nonlinear property along with the simple physical implementation has made major impacts on all areas of electronics including the information processing, communication, and power systems. On the other hand, the rectification effect in a conventional diode, although well defined under an applied voltage down to tens of mV, attenuates with a smaller signal.  The current-voltage relation eventually recovers symmetry across the zero bias; i.e., $G(V) \rightleftarrows G(-V)$ as $V\rightarrow 0$, where $G$ denotes the conductance at the channel voltage $V$.

The physical reason for the symmetric low-voltage conductance in the non-coherent regime is based on the principle of detailed equilibrium, i.e., the invariance of scattering characteristics with respect to the exchange of incident and scattered particles \cite{Hanggi2009}. In quantum mechanics, this property is attributed to the Born approximation that is often treated as a universal quality of electronic scattering \cite{Belinicher1980}.  More formally, it follows the Onsager reciprocal relation $G(V,\mathbf{H})=G(-V,-\mathbf{H})$, where $\mathbf{H}$ is the magnetic field.   The involvement of magnetic field clearly opens a door to the possibility of  breaking the conductance symmetry with respect to the voltage alone as $G(V,\mathbf{H})$ does not have to be equal to $ G(-V,\mathbf{H})$. Similarly, the correlation between the magnetic influence and the electrical transport may, in principle, lead to the asymmetry even for $V\rightarrow 0$.  Nevertheless, accessing the zero-voltage inequality remains a significant challenge despite numerous attempts \cite{Belinicher1980,Linke1999, Linke2000,Lofgren2004,Marlow2006,Forster2008,Sanchez2004,Szafran2009,Meair2012,Costache2010,Gorbatsevich2001}.
A system with high intrinsic asymmetry could be far more advantageous in realizing the desired characteristics.

The spin-orbit interaction (SOI) is thus a primary consideration.  Combined with spin polarized injection, it was shown theoretically that the SOI can cause transmission imbalance (zero-bias) in an asymmetric {\it open} structure by differentiating the passes for the spin-up and spin-down electrons in the coherent quantum mechanical regime \cite{Masuda2007}.  Even under more conventional transport conditions, spin injection at an arbitrary small voltage creates a non-equilibrium spin distribution in the non-magnetic medium. The SOI transfers spin polarization to the electronic flux, directing it according to the spin and not to the voltage polarity \cite{Silsbee2004}. This net effect of spin-dependent flux on the total current could lead to the zero-voltage conductance discontinuity if the spin polarization is conserved.
A prominent example of the spin-current conversion is the photo-voltaic effects that appears due to the spin polarization of the photo-excited electrons (e.g., see Ref.~\onlinecite{Silsbee2004} and references therein). However, the directional selectivity of conductance (and its zero-voltage discontinuity) of spin-polarized electrons has not been pursued vigorously in conventional semiconductors as the effect of scattering asymmetry is often too small to overcome the background spin-independent events. Moreover, the spin-related asymmetry decays quickly once the thermal energy becomes dominant over the spin-orbital splitting induced by the Dresselhaus or Rashba effects.  In any case, the intrinsic imbalance relying on the spin dependent properties does exist, making systems with large spin-orbital coupling potentially far more advantageous in realizing the desired characteristics.

In this regard, the recently discovered 3D topological insulators (TIs) provide a uniquely promising candidate \cite{Hasan2010}.  Unlike the traditional semiconducting 2D channels, the electron spin on  a TI surface is inherently locked to its momentum.  Thus, the electron spin polarization follows the surface current and vice versa~\cite{Hasan2010,Semenov2012,Li2014,Hong2012}.  The subsequent directional selectivity should immediately result in the ratchet effect when combined with the ferromagnetic (FM) contacts that enable spin polarized injection and collection.  Indeed, a recent measurement illustrated a difference in the TI surface resistance between two FM electrodes at finite positive/negative voltages \cite{Tian2014}.  Yet, beside being small, it is highly unclear whether the observed inequality persists at zero bias.  In the ballistic regime, the transition matrix symmetry based on charge conservation imposes the conductance reciprocity in a strictly two-terminal structure \cite{Comm}.  At the same time, theoretical calculations in a long channel (i.e., the diffusive transport) also reveal the reversible conductance as the initial net polarization quickly undergoes relaxation after a number of scattering events \cite{Burkov2010,Schwab2011}.  The desired asymmetric response clearly requires an additional consideration while not causing a significant loss of electron spin polarization.

\begin{figure}[tbp]
\includegraphics[width=8.0cm]{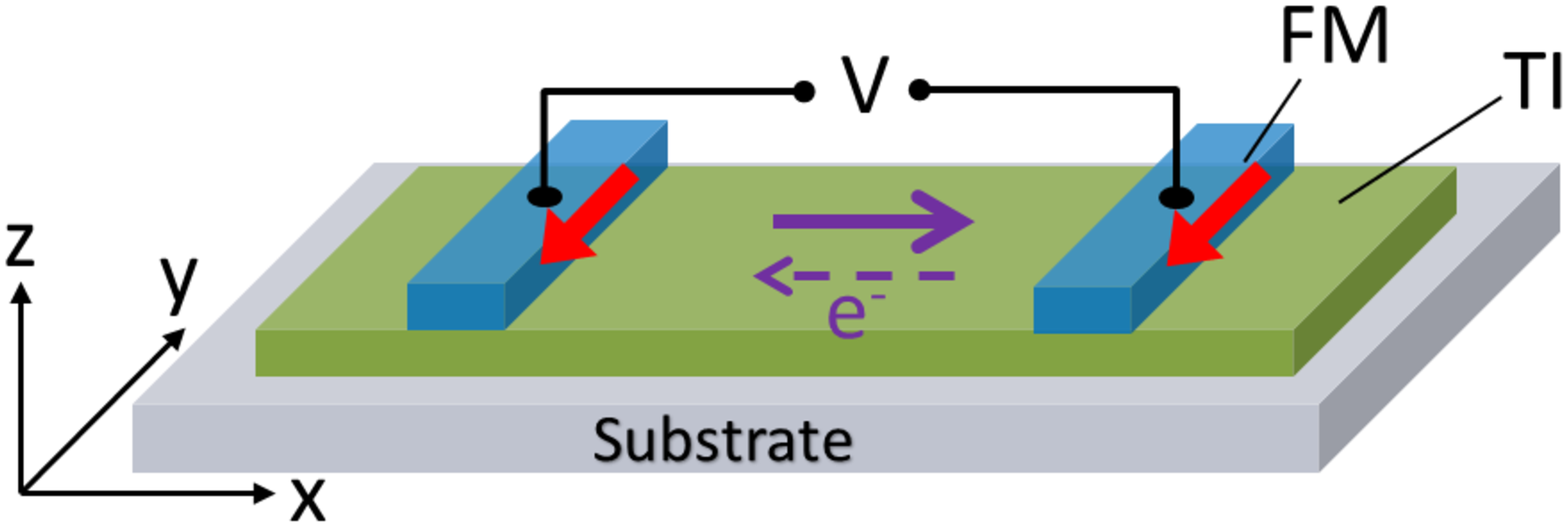}
\caption{(Color Online)
A diode-like TI structure with two magnetic electrodes on the top surface.  Thin tunnel barriers separate the TI surface and the metallic electrodes (not shown).  Two red arrows denote the magnetization direction of the magnets (FM), while $V$ stands for the applied bias. The purple arrows (thicker vs.\ thinner) indicate the preferred direction of electron flow corresponding to the specified magnetization profile.
}
\label{fig1}
\end{figure}

In the present work, we propose a solution to the long running challenge by exploiting the advantages of a TI based structure for highly asymmetric conduction at arbitrarily small forward and reverse biases.   The key variation from the conventional two-terminal assumption, as shown in Fig.~\ref{fig1}, is the finite size of the FM electrodes  in the driving direction (i.e., $x$), which are also sufficiently away from the outer edges of the TI sample (to avoid the boundary scattering).  This enables the electrons on the TI surface to establish a non-isotropic, non-equilibrium distribution following the spin injection conditions of the FM contact (with  magnetization $\mathbf{M}$ along the $y$ axis) that remains significant over the scale of spin relaxation length at an infinitesimal channel voltage.  In a sense, the TI layer may be considered as a spin sensitive scatterer that separates the electrons in the real space according to their spin momenta$-$analogous to that in the Stern-Gerlach experiment.  Thus, spin specific electron transport can be manifested in the directional conduction so long as the polarization survives the relaxation in the channel, leading potentially to $G(V\rightarrow 0^-,\mathbf{M})\neq G(V\rightarrow 0^+,\mathbf{M})$.  When the contact width is much larger than the spin relaxation length, the electron distribution on average loses the asymmetry and, hence, the polarization specific response. In fact, adoption of narrow FM contacts and subsequent electron transport in the spin-momentum interlocked channel is roughly analogous to the asymmetric open system with spin polarized injection described above \cite{Masuda2007}, even though the operating conditions are drastically different (i.e., non-coherent vs.\ coherent).

\begin{figure}[tbp]
\includegraphics[width=8.5cm]{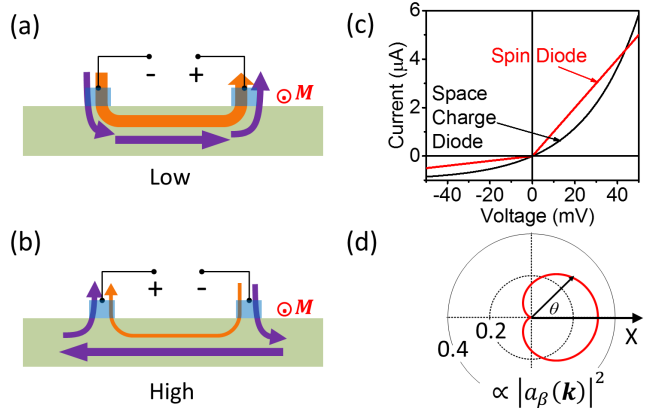}
\caption{(Color Online)
(a,b) Schematic illustration of electron flow under the forward (low resistance) and reverse (high resistance) biases, respectively.  The total current is the sum of the contributions from the the direct (orange arrows; ballistic paths) and the sequential channels (purple arrows; scattered paths). The magnetization of electrodes is assumed to be along the $-y$ axis. (c) Anticipated asymmetric $I$-$V$ characteristics near the zero bias with the rectification ratio of 10 (red).  Also shown for comparison are those of a conventional space-charge diode with a comparable current level (black).  A related description can be found in the caption for Fig.~3(b) as well.
(d) Normalized tunneling strength $\vert a_\beta(\mathbf{k})\vert^2$ between the FM electrode (with the $-y$ magnetization) and the spin-momentum interlocked TI surface states.  The area under the curve is set to 1.  The result clearly shows the imbalance between the right- and left-moving electron distribution.
}
\label{fig2}
\end{figure}

The origin of the asymmetry between the opposite bias polarities is more clearly illustrated in Figs.~\ref{fig2}(a) and \ref{fig2}(b).  Here, the light (orange) and dark (purple) arrows denote the ballistic and scattered paths, respectively.  In the low resistive case [Fig.~\ref{fig2}(a); forward bias], the electrons injected from the FM contact on average possess the TI surface momenta that are naturally aligned with the applied electrostatic force.  Hence, the conduction is primarily via one spin channel directly connecting two electrodes (for both ballistic and non-ballistic transport).  Once the sign of the applied voltage is flipped [Fig.~\ref{fig2}(b); reverse bias], however, the contribution of the direct channel involves the minority spin injection that has a significantly limited capacity and the paths involving spin/momentum altering transition must be accounted for.  The nature of this imbalance is expected to persist when the bias tends asymptotically to zero [see, for instance, Fig.~\ref{fig2}(c) as well as the related description for numerical estimation given later in the discussion].  Nevertheless the asymmetric conductance does not yield a non-zero current at zero voltage.

Our theoretical analysis predicts the presence of inequality even at infinitesimally small biases, leading to potential discontinuance in the channel conductance as the separation between the two electrodes ($L$) shrinks.  In the experimental implementation, the required dimension may not be as demanding as it first looks.  While the measurements of $\lambda$ on the TI surface suggest only about a few nm at the moment, this is due primarily to the poor sample quality.  Rather, the projected values limited by the intrinsic electron-phonon interaction could reach several $\mu$m once unintended imperfections can be minimized \cite{Glinka2013,Sobota2012,Crepaldi2013}.  Hence, achieving a high-quality material with the critical length around 100 nm appears entirely feasible.

Before examining the details, it is important to note that the finite sizes of the contacts and the distance between them complicate accurate evaluation of the electrical conductance.  More specifically, transmission via both ballistic (direct) and scattered (sequential) paths must be accounted for simultaneously even when $L \lesssim \lambda$.  As described earlier [see Fig.~\ref{fig2}(a,b)], this is because some lucky electrons injected with a "wrong" spin (thus, a momentum in the "wrong" direction) can still turn around and complete the conduction despite the restriction on the 180$^\circ$ back scattering.
Hence, the widely used treatment based on the Landauer-B\"{u}ttiker approach is not adequate, requiring further development as discussed below.

The theoretical formulation starts with the description of electrons in the FM electrodes and the TI surface channel.  In the actual realization, an atomically thin tunnel barrier may separate them (i.e., the TI and the electrodes) for highly efficient spin polarized injection as well as to avoid the modification of the TI band structure.
By considering spin-up and spin-down components explicitly, the eigenfunctions in the electrodes can be written as:
\begin{equation}\label{eq_wfm}
\psi _{s}^\mathrm{FM}=\frac{u(\mathbf{r})}{\sqrt{1+s^2}}\left(
\begin{array}{c}
-is \\
1%
\end{array}%
\right),
\end{equation}
where $s$ symbolizes the spin index and $u(\mathbf{r})$ is the spatially varying function that is determined by the structure geometry and the constituent material properties. Accordingly, $\langle\psi_s^\mathrm{FM}|\sigma_{y}|\psi_s^\mathrm{FM}\rangle=s$, where $s=\pm 1$ corresponds to $\mathbf{M}\parallel \pm\hat{y}$, respectively.
On the TI surface, a given spin polarization determines the electron momentum due to the spin-momentum interlock. The two spinor wavefunction on the TI surface is thus described by its wavevector $\mathbf{k}=(k_x,k_y)$ as:
\begin{equation}\label{eq_wfTI}
\psi _{\beta}^\mathrm{TI}=\frac{w(\mathbf{r})}{\sqrt{2}}\left(
\begin{array}{c}
\hbar v_{F}( i\beta k_{x}+k_{y})/E \\
1%
\end{array}%
\right) \exp [i(\beta k_{x}x+k_{y}y)],
\end{equation}
where $w(\mathbf{r})$ is the Bloch function (that decays exponentially from the surface along the $z$ axis) and $v_F$ is the Fermi velocity of TI surface electrons ($=4.5\times 10^5$~m/s for Bi$_2$Se$_3$).  When the electrodes are elongated along the $y$ axis as shown, the invariance in this direction can be assumed, enabling the states to be identified by the energy $E$ and transverse wavevector $k_y$. As such, $k_x$ simply becomes $\sqrt{\left( E/\hbar v_{F}\right) ^{2}-k_{y}^{2}}$ with the factor $\beta$ denoting the $+$ and $-$ signs for the rightward and leftward traveling components, respectively.
Hence, the TI electrons with a given energy and transverse wavevector can be expressed as a linear combination of the spinor eignefunctions.  While $k_y$ can take both positive and negative values, the two components ($\pm$) of the $x$ directional momentum are considered explicitly.

Then, electron transmission between the FM contact and the TI surface states can be treated via the tunneling process as in the scanning tunneling microscope \cite{Gottlieb2006}.  For instance, the stimulated surface states astride the narrow FM electrode at $x=0$ can be described as $\Psi_\mathrm{TI} = \sum_{\beta=\pm } a_{\beta}\psi _{\beta}^\mathrm{TI}\theta (\beta x)$, where $\theta (x)$ is the Heaviside step function and the coefficient $a_{\beta}$ representing electron injection in each mode simply becomes $\left\langle \psi _{\beta}^\mathrm{TI}|H_t|\psi _{s}^\mathrm{FM}\right\rangle \approx  \Delta V_b\left\langle \psi _{\beta}^\mathrm{TI}|\psi _{s}^\mathrm{FM}\right\rangle $ if a scattering potential $\Delta V_b$ is adopted as a formal representation for the tunneling Hamiltonian $H_t$ through a uniform barrier.
A straightforward calculation subsequently leads to
\begin{equation}\label{eq_wfoverlap}
a_{\beta}(\mathbf{k})=\tau_0\frac{E- s\hbar v_{F}(\beta k_{x}+ik_{y})}{2E}\,,
\end{equation}
where $\tau_0$ results from $\Delta V_b$ and the overlap integral of the spinor-independent parts, $u(\mathbf{r})$ and $w(\mathbf{r})$.  The explicit dependence of Eq.~(3) on $s$ and $\beta$ clearly illustrates the influence of spin-momentum selection; i.e., the imbalance between two transport directions when the contact is spin specific. These complex terms originating from the spin-orbit interaction make the expression non-analytic (in terms of Cauchy-Riemann conditions in the 2D parameter space) that, if persists in the TI channel, can give rise to the discontinuity in the zero-voltage conductance. As an example, the normalized directional dependence of $\vert a_\beta(\mathbf{k})\vert^2$ is plotted in Fig.~\ref{fig2}(d) for the case of $s=-1$ (i.e., with the $-y$ magnetization).   Evidently, the injected distribution favors the initial momentum along the $+x$ direction with approximately the 80:20 ratio between the majority and minority populations.  This preference is independent of the electron energy $E$.
The absorption of TI electrons by the collecting FM electrode can be handled similarly.  The only difference is a slight modification in the Heaviside step function for $\Psi_\mathrm{TI}$ as the $\beta = +$ ($-$) mode now approaches the contact from the left (right).

Following the overlap integral, the direct conduction through the ballistic paths between the two electrodes can be evaluated by treating the TI surface state as the intermediary that couples the FM states.
This results in the transmission coefficient $t_\mathrm{M}^\mathrm{M}(E)\approx -i\pi|\tau_0|^2\frac{E}{\hbar v_F}\left(1-\frac{\pi}{4}s\beta\right)$, where the proportionality to the energy originates from the density of modes on the TI surface \cite{SUPP}.
While numerical evaluation of the transmitted flux (i.e., $ | t_\mathrm{M}^\mathrm{M}|^2$) is difficult due to the uncertainty in $\tau_0$, it is outright obvious that the relative ratio of the conductance can be obtained as the polarity of the magnetization and/or the driving bias change.
Assuming both electrode are magnetized along the $-{y}$ axis ($s=-1$), for instance, a switch from the forward to reverse bias ($\beta = + \rightarrow - $) yields conductance variation as large as 70 to 1 [more precisely, $(1 + \pi/4)^2 : (1 - \pi/4)^2$].  For convenience, this factor illustrating the dependence on the magnetization and bias polarities is defined as%
\begin{equation}\label{eq_Tmm}
T_\mathrm{M}^\mathrm{M}=\frac{\left(1 - s\beta \pi/4\right)^2}{2\left(1+\pi^2/16\right)} .
\end{equation}%
The denominator normalizes the directional contribution to the sum of both spin states that corresponds to the case of nonmagnetic electrodes.  Hence, the actual flux can be obtained formally from the product of $T_\mathrm{M}^\mathrm{M}$  and the spin-independent, ballistic path conductance (denoted as $G_{\mathrm{M}}^{\mathrm{M}}$).
Following the schematic given in Fig.~\ref{fig1}, we set $s= -1$ in the ensuing discussion.  The case for the $+y$ magnetization (i.e., $s=+1$) can be evaluated similarly.

Along with the ballistic conduction between two FM electrodes, it is necessary to account for the role of the sequential channel involving the scattered paths even when $L \lesssim \lambda$.  Due to the finite width of the contact region, the electrons initially injected with the momenta moving away from the collecting contact can turn around after a few scattering events and provide a non-zero contribution.  It is particularly relevant if the conduction is in the high resistive (i.e., reverse biased) state as indicated in Fig.~\ref{fig2}(b).  When $L$ is large, on the other hand, the non-ballistic current becomes dominant and the channel conductance converges to the spin-independent diffusive limit.
Unfortunately, a rigorous treatment of this complex transport problem is very difficult to attain.  Instead, we approach it by estimating the effective length of the non-ballistic paths where the scattering events are modeled as elastic processes.
Typically, the additional distance that an electron injected into the "wrong" spin state (thus, the momentum) travels away before experiencing the turn around is of the order of the mean free path $\lambda$. Given that the back scattering on the TI surface is suppressed, the average detour $q \lambda$ is somewhat longer than $\lambda$
with a value around $q \sim 4$ following a simple analysis.
A similar process is also encountered in the absorbing electrode, causing additional differences in the paths. Of the two conduction conditions,  the high resistive state is again expected to experience stronger reflection  due to the spin mismatch and, thus, a longer path on the collecting end [see Fig.~\ref{fig2}(b)].  We neglect this extra spin (or polarity) selective contribution as it is not crucial in the main focus of the investigation (i.e., demonstrating the nonreciprocal conduction).  The resulting approximation tends to underestimate the asymmetry ratio; hence, a rather conservative assumption.

Then, the non-ballistic surface channel conductance can be given as shown on the right side of the expression \cite{SUPP}
\begin{equation}\label{eq_GTI}
G_\mathrm{TI}^L T_\mathrm{TI}= G_\mathrm{TI}^L \times \left\{
\begin{array}{lr}
p +  (1-p)\frac{1}{(1+q\lambda /L)}, ~~~&\mbox{forward}\\
(1-p) + p \frac{1}{(1+q\lambda /L)}, ~~~&\mbox{reverse}
\end{array}
\right.
\end{equation}
Here, $G_\mathrm{TI}^L$ represents the (bulk) TI surface conductance of length $L$ (hence, the conductivity of $ L G_\mathrm{TI}^L$).  In addition, the factors $p$ ($\approx 0.8$) and $1-p$ account for the uneven distribution of the injected electrons between the $+k_x$ and $-k_x$ components as discussed earlier in Fig.~\ref{fig2}(d). The portion grouped in the curly parenthesis is subsequently defined as $T_\mathrm{TI}$, highlighting the normalized spin/polarity-dependent component. To obtain the contact-to-contact transport characteristics, the (sequential) tunnel resistance  between the electrode and the TI ($1/G_\mathrm{TI}^\mathrm{M}$) must also be considered.  Hence, the total resistance in this case becomes
${1}/{G_\mathrm{TI}^L T_\mathrm{TI}}+ {2}/{G_\mathrm{TI}^\mathrm{M}}$.

In a steady state with a nonzero bias $V$, the total current is obtained by summing the contributions from both the ballistic and non-ballistic transport. Thus, the resulting total conductance [$ = I(V)/V$] follows the form:
\begin{equation}\label{eq_Gt}
G_\mathrm{tot}(V)=\left[ \exp^{- L/ \lambda} G_\mathrm{M}^\mathrm{M} T_\mathrm{M}^\mathrm{M} +
\frac{ G_\mathrm{TI}^L G_{\mathrm{TI}}^\mathrm{M} T_\mathrm{TI}}{G_{\mathrm{TI}}^\mathrm{M}+2
G_\mathrm{TI}^L T_\mathrm{TI}} \right].
\end{equation}
As shown, the exponential decay is considered explicitly in the ballistic contribution (the first term), while the second accounts for the diffusive conduction described directly above. Under a small voltage, the $I$-$V$ curve is expected to be linear in each of the forward and reverse bias regions but with different slopes [i.e., $G_\mathrm{tot}(V) \neq G_\mathrm{tot}(-V)$; see also the illustration in Fig.~2(c)].  Such a possibility is evident from the dependence of Eqs.~(\ref{eq_Tmm}) and (\ref{eq_GTI}) on the relevant parameters.

\begin{figure}[tbp]
\includegraphics[width=8.5cm]{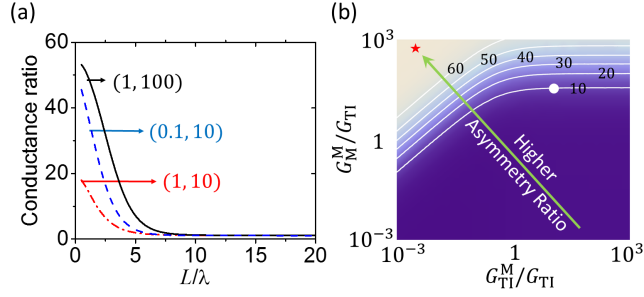}
\caption{(Color Online) (a) Conductance ratio between the forward and reverse biases of infinitesimal amplitude (i.e., the conductance asymmetry or discontinuity) as a function of the normalized channel length $L/\lambda$ for three different combinations of ($G_{\mathrm{TI}}^\mathrm{M}/G_\mathrm{TI}$, $ G_\mathrm{M}^\mathrm{M}/G_\mathrm{TI}$).  (b) Contour plot of the conductance asymmetry ratio in the $G_{\mathrm{TI}}^\mathrm{M}/G_\mathrm{TI}$-$ G_\mathrm{M}^\mathrm{M}/G_\mathrm{TI}$ parameter space for $L = 3 \lambda$.   The star indicates the limit of approx.\ 70 that is set by the purely ballistic transport.  The numbers denote the calculated asymmetry values.  The lighter (darker) color signifies the higher (lower) ratio. The white dot indicates the parameters used for the $I$-$V$ calculation (spin diode) shown in Fig.~\ref{fig2}(c).
}
\label{fig3}
\end{figure}

For the quantitative results, a parametric study is carried out as the numerical values of $G_\mathrm{M}^\mathrm{M}$, $G_{\mathrm{TI}}^\mathrm{M}$, and $ G_\mathrm{TI}^L$ are not established. To simplify the analysis, the surface conductance $ G_\mathrm{TI}$ defined for a fixed length $\lambda$ (electron mean free path) is used hereinafter; i.e.,
$
G_\mathrm{TI}^L \Rightarrow \frac{\lambda}{L}G_\mathrm{TI}
$
in Eqs.~(\ref{eq_GTI}) and (\ref{eq_Gt}) showing explicitly the dependence on $L$. In this regard, the channel dimension is described in units of $\lambda $ as well.  Figure~\ref{fig3}(a) shows the conductance ratios between the forward and reverse biases of infinitesimal amplitude for three different combinations of $G_{\mathrm{TI}}^\mathrm{M}/G_\mathrm{TI}$ and $G_\mathrm{M}^\mathrm{M}/G_\mathrm{TI}$.  The rapid decay with the increasing $L$ conforms to the relaxation of the spin selective injection through the interaction with the environment.  Once $L\gg \lambda$, the transport properties converge to the drift-diffusion equations for all cases under consideration \cite{Burkov2010}.  In the other extreme, the ballistic conduction dominates and a large asymmetry over the factor of 10 can be clearly achieved as desired. For the strong rachet effect, the calculation results also indicate the need for a large $G_\mathrm{M}^\mathrm{M}$ and/or a small $G_{\mathrm{TI}}^\mathrm{M}$ in reference to $ G_\mathrm{TI}$.

The impact of $G_{\mathrm{TI}}^\mathrm{M}/G_\mathrm{TI}$ and $G_\mathrm{M}^\mathrm{M}/G_\mathrm{TI}$ is analyzed in  greater detail  with $L$ fixed at $3 \lambda$ in Fig.~\ref{fig3}(b).  The star on the upper left corner indicates the limit of approx.\ 70 set by $T_\mathrm{M}^\mathrm{M}$.  Evidently, the ratio approaches to the maximum when the conductance is dominated by the ballistic transport (i.e., a large $G_\mathrm{M}^\mathrm{M}/G_\mathrm{TI}$).  As for the dependence on $G_{\mathrm{TI}}^\mathrm{M}/G_\mathrm{TI}$, the asymmetry ratio is not a monotonously decreasing function despite the initial impression.  At the low limit, the large tunnel resistance between the electrode and the TI effectively blocks the sequential conduction involving the scattered paths, leaving the ratio to follow that of the direct conduction via ballistic paths.  This diminishes any contribution from $T_\mathrm{TI}$ [Eq.~(\ref{eq_GTI})] for the asymmetry.  Once $G_\mathrm{TI}^\mathrm{M}$ increases, the scattered paths start to draw the current and the ratio decreases rapidly as expected (i.e., a more significant role of the diffusive regime with a smaller ratio).
When $G_\mathrm{TI}^\mathrm{M}$ becomes even larger toward the high limit, the asymmetry ratio deviates somewhat from the simple decay with the value determined by the interplay between $G_{\mathrm{M}}^\mathrm{M}T_{\mathrm{M}}^\mathrm{M}$ and $G_{\mathrm{TI}}T_\mathrm{TI}$.
One potential complication is that $G_{\mathrm{TI}}^\mathrm{M}$ and $G_\mathrm{M}^\mathrm{M}$ may not be controlled independently as both of them are related to the tunneling characteristics. Nonetheless, a thinner tunnel barrier is preferred since ensuring a large contribution from the ballistic paths is more crucial for high asymmetry. On a side note, the white dot in Fig.~3(b) denotes the conditions under which the $I$-$V$ characteristics in Fig.~\ref{fig2}(c) are obtained.  The piecewise linear relation with a sizable difference in the forward/reverse slopes (a factor of 10) provides distinctively unique features compared to the conventional diode with a similar current scale.

In summary, the asymmetric conductance between two narrow magnetic electrodes on a TI surface is predicated.
This phenomenon requires that the distance between electrodes be comparable to the electron mean free path and the electrodes far from the TI boundaries to minimize the relaxation.  The spin selective effect of the magnetic electrodes coupled with the intrinsic spin-momentum interlock causes the conductance imbalance which is far more significant than that based on the structural asymmetry. Our theoretical estimate indicates a clearly measurable conductance imbalance between the forward and reverse biases even at infinitesimally small voltages.  A key for experimental verification at room temperature is the improved sample quality as the characteristic dimension is limited by $\lambda$.

This work was supported, in part, by FAME (one of six centers of STARnet, a SRC program sponsored by MARCO and DARPA) and US Army Research office.

\clearpage

\clearpage

\clearpage

\clearpage

\end{document}


\title{Supplementary: Conductance Discontinuity on the Surface of a Topological Insulator with Magnetic Electrodes}

\author{Xiaopeng Duan}
\affiliation{Department of Electrical and Computer Engineering, North Carolina State University, Raleigh, NC 27695, USA}

\author{Xi-Lai Li}
\affiliation{Department of Electrical and Computer Engineering, North Carolina
State University, Raleigh, NC 27695, USA}

\author{Yuriy G. Semenov}
\affiliation{Department of Electrical and Computer Engineering, North Carolina State University, Raleigh, NC 27695, USA}

\author{Ki Wook Kim}\email{kwk@ncsu.edu}
\affiliation{Department of Electrical and Computer Engineering, North Carolina State University, Raleigh, NC 27695, USA}
\affiliation{Department of Physics, North Carolina State University, Raleigh, NC 27695, USA}

\maketitle

\section{Calculation of ballistic transmission coefficient between electrodes}
With the only assumption of spin conservation, the coupling between the electrodes and the TI surface [i.e., $\langle \psi^\mathrm{FM}(E,s)\vert \Delta V_b \vert \psi^\mathrm{TI}_{\beta}(k_x,k_y)\rangle$] follows the spin selective rule, which results in the coupling coefficient dependent on the TI electron momentum as shown by Eq.~(3) in the main paper. Here, it should be noted that the states in the electrode is described by the energy $E$ and the spin polarization $s$, while that in the TI is by the momentum according to the linear dispersion relation and the spin-momentum interlock.
The overall physical picture for the direct ballistic coupling between two electrodes can be constructed by treating the TI surface states as the intermediary.  Then, by taking advantage of the
overlap integral through the tunnel barrier mentioned above, a formal expression for this coupling coefficient between electrodes can be written as:
\begin{equation}\label{eq_tmm0}
t_\mathrm{M}^\mathrm{M}(E)=\sum_{k_x^\prime}
~\sum_{k_y^\prime}
\frac{\langle \psi^\mathrm{FM}_{R}(E,s)\vert \Delta V_b \vert \psi^\mathrm{TI}_{\beta}(k_x^\prime,k_y^{\prime})\rangle\langle \psi^\mathrm{TI}_{\beta}(k_x^\prime,k_y^{\prime})\vert  \Delta V_b \vert  \psi^\mathrm{FM}_{L}(E,s)\rangle e^{i\beta k_x^\prime L}}{E-E'+i0} ,
\end{equation}
where the subscripts for the FM states represent the electrode on the left ($L$) or the right ($R$) and $(\beta k_x^\prime,k_y^\prime)$ is the momentum of the TI surface state that couples to the electrodes. As in the main paper, we set $k_x^\prime$ a positive value and use $\beta$ to denote the direction of electron flow. Given a width $W$ for the channel, $k_y^\prime$ is then quantized as $k_y^\prime=m\pi/W$, where $m=0,\pm 1, \pm 2, \cdots$. If the electron has an energy $E^\prime=\hbar v_F \sqrt{{k_x^\prime}^2+{k_y^\prime}^2}$, the upper bound for $|m|$ is determined accordingly by $E^\prime$; i.e., $|m|\leq E^\prime W/\pi\hbar v_F$.  When the width is sufficiently large such that the quantization step $\pi/W$ is negligible compared to $E^\prime/\hbar v_F$, the summation over all possible $k_y^\prime$ can be substituted by an integral. The same applies to the ${x}$ direction as well. Indeed, we assume that the ${x}$ direction is unconstrained, so $k_x^\prime$ essentially becomes continuous. In addition, it is more suitable to average the result over the system dimension.  We adopt the change of notation to let $t_\mathrm{M}^\mathrm{M}$  denote such an averaged value hereinafter, i.e.,
\begin{equation}\label{eq_tmm}
\begin{split}
t_\mathrm{M}^\mathrm{M}(E)=&\int_{-\infty}^{\infty}
~\int_{-\infty}^{\infty}
\frac{\langle \psi^\mathrm{FM}_{R}(E)\vert \Delta V_b \vert \psi^\mathrm{TI}_{\beta}(k_x^\prime,k_y^{\prime})\rangle\langle \psi^\mathrm{TI}_{\beta}(k_x^\prime,k_y^{\prime})\vert  \Delta V_b \vert  \psi^\mathrm{FM}_{L}(E)\rangle}{E-E'+i0} \frac{k_x^\prime}{\sqrt{{k_x^\prime}^2+{k_y^\prime}^2}} dk_x^\prime dk_y^\prime
\\
=&\int_{-\infty}^{\infty}
~\int_{-\pi/2}^{\pi/2}
\frac{\langle \psi^\mathrm{FM}_{R}(E)\vert \Delta V_b \vert \psi^\mathrm{TI}_{\beta}(E^\prime,\theta)\rangle\langle \psi^\mathrm{TI}_{\beta}(E^\prime,\theta)\vert  \Delta V_b \vert  \psi^\mathrm{FM}_{L}(E)\rangle}{E-E'+i0} \cos\theta \frac{E^\prime}{\hbar v_F} d\theta dE^\prime .
\end{split}
\end{equation}
This is a quasi-1D treatment that is valid given the invariance over ${y}$.  The second expression indicates the change of variables according to $dk_x^\prime dk_y^\prime=\frac{E^\prime}{\hbar v_F}dE^\prime d\theta$ (i.e., the $E$-$k$ relation). In addition, the phase factor $e^{ik_x^\prime L}$ is dropped from the equation as it is for the interference of electrons with different wavevectors.  The interference phenomenon requires very stringent conditions that are generally not supported in the realistic systems at room temperature.  The $\cos\theta$ factor, or the $\frac{k_x^\prime}{\sqrt{{k_x^\prime}^2+{k_y^\prime}^2}}$ term, explicitly accounts for the projection along the channel direction. The integral over the energy can be carried out by using Cauchy's residue theory. This is intuitively expected as the resonant case with $E^\prime=E$ dominates the contribution. Thus, Eq.~(\ref{eq_tmm}) becomes:
\begin{equation}\label{eq_tmm1}
t_\mathrm{M}^\mathrm{M}(E)=-i\pi\int_{-\pi/2}^{\pi/2}
\langle \psi^\mathrm{FM}_{R}(E)\vert \Delta V_b \vert \psi^\mathrm{TI}_{\beta}(E,\theta)\rangle\langle \psi^\mathrm{TI}_{\beta}(E,\theta)\vert  \Delta V_b \vert  \psi^\mathrm{FM}_{L}(E)\rangle \cos\theta d\theta .
\end{equation}
Combined with the overlap integral given in Eq.~(3) of the main paper, the coupling between the two electrodes through the TI surface states is finally given as (with $k_x^\prime=E \cos\theta/\hbar v_F$ and $k_y^\prime=E \sin\theta/\hbar v_F$):
\begin{equation}\label{eq_tmm2}
t_\mathrm{M}^\mathrm{M}(E)=-i\pi\int_{-\pi/2}^{\pi/2}
\vert \tau_0\frac{E-s\hbar v_F(\beta k_x^\prime+ik_y^\prime)}{2E}\vert^2 \cos\theta
d\theta
=-i\pi|\tau_0|^2\frac{E}{\hbar v_F}\left(1-\frac{\pi}{4}s\beta\right) .
\end{equation}

\section{Estimation of diffusive channel conductance}
As described in the main text, the non-ballistic diffusive conduction can come from the electrons injected with both the "correct" spin (moving toward the collecting contact) and the "wrong" spin (moving away from the collecting contact).  The electrons in the first group ("correct" spin) travel the distance of $L$ (i.e., the separation between the two contacts) to reach the collecting electrode, while those in the second group ("wrong" spin) experience longer paths ($L + q \lambda$) since they have to first turn around via scattering events ($\lambda$ being the electron mean free path).  When the conductivity of the channel is given by $LG_\mathrm{TI}^L$ in the diffusive regime (as specified in the main text), the conductance of the former and the latter paths becomes $G_\mathrm{TI}^L$ and $G_\mathrm{TI}^L {L}/{(L + q \lambda)}$, respectively, reflecting the difference in the path lengths.  Then, by treating the probability of an electron injected with the "correct" spin to be $p^\prime$ and that with the "wrong" spin $1-p^\prime$, the diffusive channel conductance can be written approximately as:
\begin{equation}
G_\mathrm{diff} = p^\prime G_\mathrm{TI}^L + (1 - p^\prime)  \frac{L}{L + q \lambda} G_\mathrm{TI}^L \,.
\end{equation}
When the structure is forward biased, $p^\prime$ corresponds to $p$ ($\approx 0.8$) in the main text. For the reverse biased case, it is $1-p$ that matches with $p^\prime$ instead.  This leads to Eq.~(5) of the main text.  Actually, the latter expression adopts an additional parameter $ T_\mathrm{TI}$ that is normalized by $G_\mathrm{TI}^L $ to highlight the spin/polarity dependence (i.e., $ T_\mathrm{TI} = G_\mathrm{diff}/G_\mathrm{TI}^L $).